\documentclass[aps,superscriptaddress,prc,eprint,nofootinbib,showkeys]{revtex4-2}
\pdfoutput=1
\usepackage{graphicx}
\usepackage{dcolumn}
\usepackage{bm}

\usepackage{amsmath,amsfonts,amsthm,amssymb}
\usepackage{graphics}
\usepackage{cancel}
\usepackage{multirow}
\usepackage{longtable}
\usepackage{color}
\usepackage[normalem]{ulem} 

\usepackage{amssymb}

\usepackage{graphicx}
\usepackage{amsmath}
\usepackage{subfigure}
\usepackage{bbold}
\usepackage{yfonts}
\usepackage[colorlinks=true,linktocpage=true,linkcolor=blue,citecolor=blue]{hyperref}
\usepackage{placeins}
\usepackage{bm} 
\usepackage{nicefrac}
\usepackage{slashed}
\usepackage{marginnote}

\newcommand{\ab}[1]{\left\langle#1\right\rangle}

\begin{document}
	
	\preprint{APS/123-QED}

	\title{Relativistic spin hydrodynamics with momentum and spin-dependent relaxation time}
	\author{Samapan Bhadury}
	\email{samapan.bhadury@uj.edu.pl}
	\affiliation{Institute of Theoretical Physics, Jagiellonian University ul. St.\L ojasiewicza 11, 30-348 Krakow, Poland}
	
    \date{\today}
	
    \begin{abstract}
        Using the extended relaxation time approximation (ERTA) along with the theory of semi-classical spin, we develop a framework of relativistic dissipative spin hydrodynamics such that the relaxation time can depend on the momenta and spin of the constituent spin-1/2 particles. We also consider a general definition of the fluid four-velocity allowing the theory to be valid in a general frame and matching conditions. Consequently, we construct the frame-invariant bulk, shear, particle diffusion, and spin transport coefficients, showing that the evolution of fluid remains unaffected by spin in the limit of small polarization as was the case where the relaxation time was independent of spin or momentum.
    \end{abstract}
         
    \date{\today}


\maketitle


\section{Introduction}
\label{intro}

Spin polarization of hadrons and vector mesons in heavy-ion collision (HIC) experiments has become a new observable of high interest in recent years \cite{STAR:2017ckg, STAR:2018gyt, ALICE:2019aid, Liang:2004ph, Liang:2004xn, Liang:2004xn, Becattini:2009wh, Becattini:2013fla, Li:2017slc, Wang:2017jpl, Sun:2017xhx, Becattini:2017gcx, Florkowski:2018ahw, Yang:2018lew, STAR:2018pps, STAR:2018fqv, Niida:2018hfw, Florkowski:2018fap, Florkowski:2019qdp, STAR:2019erd, ALICE:2019onw, Singha:2020qns, Chen:2020pty, Becattini:2020sww, Singh:2020rht, STAR:2020xbm, Goncalves:2021ziy, ALICE:2021pzu, STAR:2021beb, Singh:2021yba, Florkowski:2021wvk, Mohanty:2021vbt, STAR:2022fan, STAR:2023eck, Kumar:2023ghs, Sheng:2023urn, Gevorkyan:2023tzb, Fang:2023bbw, Kumar:2023ojl, Lv:2024uev, Sheng:2024kgg, Li:2024qae, Palermo:2024tza, Yi:2024kwu} as several experimental observations remain unexplained in spite of the advances made on the theoretical side. These include the famous ``spin sign puzzle" associated with the quantitative as well as qualitative mismatch between the theoretical prediction and experimental observation of longitudinal polarization of $\Lambda$-hyperons. The failure of quantitative prediction for the global polarization of the vector mesons is another unresolved issue. In the last couple of years, there have been some proposals to explain the former through the inclusion of ``shear-induced polarization" (SIP) \cite{Fu:2021pok, Becattini:2021suc, Becattini:2021iol} but these theories are not free of ambiguities. On the other hand, alternative theories, such as the inclusion of dissipative effects has also explained the observation \cite{Banerjee:2024xnd}. While the exact explanation for the longitudinal spin polarization is not yet agreed upon, the theoretical works over the years have shed light on some other important aspects associated with the dynamics of spin within the relativistic fluid. For example, several recent works indicate that the timescale for spin relaxation should be greater than the timescale associated with other transport properties \cite{Kapusta:2019sad, Ayala:2019iin, Ayala:2020ndx, Hidaka:2023oze, Wagner:2024fhf, Banerjee:2024xnd}. These works however have treated the relaxation times as free parameters and do not consider how the spin may affect the corresponding timescales. In this work, we develop a theory of relativistic spin hydrodynamics where the relaxation times can depend on the spin as well as on the momenta of the constituent particles. For this purpose, we will use the recently developed extended relaxation time approximation. A further advantage of this framework is that we can construct the theory in a general frame without specifying the frame and matching conditions a priori\footnote{However, it should be noted that this generality of the frame and matching conditions only implies that we are not forced to have a specific choice to ensure the conservation laws. We still need to specify the choice to have the explicit expressions of the transport coefficients~\cite{Kovtun:2019hdm}.}. Therefore, we also specify the set of frames and matching invariant transport coefficients.

In the context of transport properties of the matter formed in HIC experiments, relativistic dissipative hydrodynamics have played a crucial role. Hydrodynamics is a macroscopic theory describing the dynamics of the long-wavelength, low-frequency modes. The system produced by HIC sits at a very interesting length scale where hydrodynamics is applicable yet the quantum effects are non-negligible. This property of the system, due to the developing interest in the spin dynamics of the QCD matter, allowed the successful incorporation of spin in the theory of relativistic hydrodynamics resulting in the formulation of relativistic ideal spin hydrodynamics from Lagrangian theory \cite{Montenegro:2017rbu, Montenegro:2017lvf}, as well as kinetic theory \cite{Florkowski:2017ruc, Florkowski:2017dyn}. This was later extended to dissipative case \cite{Montenegro:2018bcf, Bhadury:2020puc, Shi:2020htn, Bhadury:2020cop, Hu:2021pwh} and the effect of external fields was also considered \cite{Bhadury:2022ulr, Buzzegoli:2022qrr}. On the other hand, in a series of papers, the effect of non-local collisions was explored \cite{Weickgenannt:2020aaf, Weickgenannt:2021cuo, Weickgenannt:2022qvh, Wagner:2022amr}. In an alternative approach, the connection of spin with torsion was also studied \cite{Hongo:2021ona, Gallegos:2022jow}. These developments eventually led to the studies of spin modes and associated spin relaxation timescales \cite{Hongo:2021ona, Hongo:2022izs, Ambrus:2022yzz, Hu:2022xjn, Sarwar:2022yzs, Daher:2022wzf, Xie:2023gbo, Weickgenannt:2023btk, Wagner:2024fhf, Ren:2024pur}. Although there has been some efforts in obtaining the Kubo formulas of spin hydrodynamics \cite{Hu:2021lnx, Tiwari:2024trl}, the computation of the spin transport based on some underlying microscopic theory has been very limited \cite{Bhadury:2020cop, Hu:2021pwh, Bhadury:2022ulr}. These works utilized the relativistic kinetic theory for the formulation of relativistic spin hydrodynamics.

In recent years, relativistic kinetic theory, as a basis of relativistic hydrodynamics, has seen several new developments, which have led to the determination of the transport coefficients with the help of the Boltzmann transport equation. While the left-hand side of the relativistic transport equation describes the evolution of the distribution function, the linearized collision operator on the right-hand side contains the details of the collision mechanism of the system. The relaxation time appearing in these collision kernels holds the information regarding the cross-sections of the collisions and hence can, in general, be a function of spacetime as well as the particle's properties such as momentum and spin. However, until recently the theories of relativistic hydrodynamics built from kinetic theories consisted of relaxation times that were independent of such particle properties. Making the relaxation time, dependent on the particle's momenta led to inconsistencies in the formulation of relativistic hydrodynamics with the relaxation time approximation proposed by Anderson-Witting \cite{Anderson:1974nyl}. Presently, several proposals have been made to allow momentum dependence of the relaxation time \cite{Rocha:2021zcw, Dash:2021ibx, Biswas:2022cla}. In the present work, we make further use of the newly developed ERTA to allow the relaxation time to be dependent on the spin of the particles as well. Recently, the transseries structure of ERTA was explored \cite{Kamata:2022ola} and this collision kernel was also used to formulate relativistic magnetohydrodynamics \cite{Singh:2024leo}. Taking advantage of the properties of this collision kernel, we carry out our calculations in a general frame with arbitrary matching conditions and determine the frame-invariant transport coefficients. This allowed us to obtain the expression of the spin transport in frame-invariant form, for the first time. Thus the present work not only shows the approach to include spin within the ERTA framework, but also shows how to compute the transport coefficient independent of frame and matching conditions within ERTA for the first time as well.

The article is organized as follows: Section~\ref{sec:RSH} introduces the theory of relativistic spin hydrodynamics under an arbitrary choice of fluid frame and matching conditions. Section~\ref{sec:TP} describes the kinetic theory framework, used for the construction of relativistic spin hydrodynamics, shows the method of solving the Boltzmann equation with the extended relaxation time approximation that incorporates spin degrees of freedom. The latter parts of the section focus on the determination of transport coefficients, which are also obtained in a frame-invariant manner by using entropy production, and finally the Section~\ref{sec:TP} is concluded by discussing some special cases of spin dependence of the relaxation time. In the Section~\ref{sec:S&O} we provide a summary and outlook of the present work.

\textit{Notation and Conventions:} In this article we have adopted the natural units i.e. $k_B = \hbar = c = 1$. We use the mostly positive metric i.e. $g^{\mu\nu} = diag(1,-1,-1,-1)$ and for the Levi-Civita tensor we have considered the convention, $\epsilon^{0123} = - \epsilon_{0123} = 1$. The spin four-vector ($s^\alpha$) and the spin angular momentum ($s^{\alpha\beta}$) are related to each other as, $s^\alpha = \epsilon^{\alpha\beta\mu\nu} p_\beta s_{\mu\nu}/2m$ and, $s^{\alpha\beta} = \epsilon^{\alpha\beta\mu\nu} p_\mu s_{\nu}/m$. For scalar products of vectors sometimes we will use $A \cdot B \equiv A_\mu B^\mu$ whereas for scalar produces of tensors we will use, $A:B \equiv A_{\mu\nu} B^{\mu\nu}$. For the Lorentz indices, we will assume the use of usual Einstein summation convention.

\section{Relativistic Spin Hydrodynamics}
\label{sec:RSH}

The theory of relativistic spin hydrodynamics involves the conservation laws of energy-momentum tensor ($T^{\mu\nu}$), particle four-current ($N^{\mu}$), and total angular momentum tensor ($J^{\lambda,\mu\nu}$). For a fluid constituted of particles with non-zero internal spin degrees of freedom, the last quantity can be split into orbital and spin parts as, $J^{\lambda,\mu\nu} = x^\mu T^{\lambda\nu} - x^\nu T^{\lambda\mu} + S^{\lambda,\mu\nu}$ where $x^\mu$ denotes the spacetime four-vector and $S^{\lambda,\mu\nu}$ denotes the spin tensor. This splitting of angular momentum is not free of ambiguity and gives rise to the now well-known problem associated with pseudogauge transformation \cite{Hehl:1976vr}. This has led to some interesting investigations recently \cite{Dey:2023hft, Buzzegoli:2024mra}. In the present work, we will not attempt to address this problem, rather we work with the specific choice of Groot-Leeuwen-Weert (GLW) pseudogauge \cite{DeGroot:1980dk} where the relativistic kinetic theory with spin is well established \cite{Florkowski:2018fap}. Then we can express these conserved currents of an out-of-equilibrium relativistic fluid, in terms of a single-particle phase-space distribution function with an extended phase-space i.e. $f(x,p,s)$ as,
\begin{subequations}
    \begin{align}
        N^\mu &= \left(n_0 + \delta n\right) u^\mu + n^\mu = \int dP dS\, p^\mu \left(f - \Bar{f}\right), \label{N-def}\\
        T^{\mu\nu} &= \left(\varepsilon_0 + \delta \varepsilon\right) u^\mu - \left(P_0 + \delta P\right) \Delta^{\mu\nu} + 2 h^{(\mu} u^{\nu)} + \pi^{\mu\nu}
        = \int dP dS\, p^\mu p^\nu \left(f + \Bar{f}\right), \label{T-def}\\
        S^{\lambda,\mu\nu} &= S_0^{\lambda,\mu\nu} + \delta S^{\lambda,\mu\nu} = \int dP dS\, p^\lambda s^{\mu\nu} \left(f + \Bar{f}\right), \label{S-def}
    \end{align}
\end{subequations}
where $dP = \frac{g d^3p}{\left(2\pi\right)^3 p^0}$ is the Lorentz invariant momentum space integral measure with $g$ being the degeneracy factor (excluding spin degrees of freedom), $dS = \frac{m}{\pi \mathfrak{s}} d^4 s \delta\left(s\cdot s + \mathfrak{s}^2\right) \delta \left(p\cdot s\right)$ is the spin integral measure with $\mathfrak{s}^2$ being the eigenvalue of the Casimir operator which for the present case of spin-1/2 particles takes the value of $\frac{3}{4}$, $f$ and $\Bar{f}$ denotes the particle and anti-particle phase-space distribution functions respectively in the extended phase-space that includes spin along with spacetime and momentum. In Eqs.~\eqref{N-def}-\eqref{S-def}, $n_0, \varepsilon_0, P_0, S_0^{\lambda,\mu\nu}$ denote the equilibrium number density, energy density, pressure and spin current respectively\footnote{Throughout this article, we use a `$0$' in the subscript to denote the equilibrium variables and a $\delta$ in front to denote their deviation from equilibrium values. Then for some out-of-equilibrium quantity $X$ we can write $X = X_0 + \delta X$.} whereas we use $\delta n, \delta \varepsilon, \delta P$ to denote the out-of-equilibrium corrections to number density, energy density, and pressure respectively. We have also used $n^\mu, h^\mu, \pi^{\mu\nu}$ and $\delta S^{\lambda,\mu\nu}$ to denote the particle diffusion current, heat diffusion current, shear viscous tensor and off-equilibrium spin current respectively. The tensor decomposition in Eqs.~\eqref{N-def}-\eqref{S-def} is carried out with the help of the timelike four-vector, $u^\mu$ which has been normalized to unity i.e. $u_\mu u^\mu = 1$ and the orthogonal projection operator, $\Delta^{\mu\nu} = g^{\mu\nu} - u^\mu u^\nu$. In writing Eqs.~\eqref{N-def}-\eqref{S-def} we have not chosen any specific definition of the fluid four-velocity, $u^\mu$ i.e. we have not imposed any frame and matching condition, and consequently, the results of this work will be for a general fluid frame with arbitrary matching conditions unless otherwise specified.

The conservation of the particle four-current ($\partial_\mu N^\mu = 0$), the energy-momentum tensor ($\partial_\mu T^{\mu\nu} = 0$) and, the spin current $(\partial_\lambda S^{\lambda,\mu\nu} = 0$) leads to the following hydrodynamic equations,
\begin{subequations}
    \begin{align}
        &\Dot{n}_0 + \delta \Dot{n} + \left(n_0 + \delta n\right) \theta + \left(\partial_\mu n^\mu\right) = 0, \label{Heq1}\\
        &\Dot{\varepsilon}_0 + \delta \Dot{\varepsilon} + \left(\varepsilon_0 + P_0 + \delta \varepsilon + \delta P\right) \theta + \left(\partial \cdot h\right) - h^\nu \Dot{u}_\nu - \pi^{\mu\nu} \sigma_{\mu\nu} = 0, \label{Heq2} \\
        &\left(\varepsilon_0 + P_0 + \delta \varepsilon + \delta P\right) \Dot{u}^\alpha - \nabla^\alpha \left(P_0 + \delta P\right) + h^\mu \left(\nabla_\mu u^\alpha\right) + \Delta^\alpha_\nu \Dot{h}^\nu + h^\alpha \theta + \Delta^\alpha_\nu \partial_\mu \pi^{\mu\nu} = 0, \label{Heq3}\\
        &\partial_\lambda S_0^{\lambda,\mu\nu} + \partial_\lambda \delta S^{\lambda,\mu\nu} = 0. \label{Heq4}
    \end{align}
\end{subequations}
where we have used the notations $\Dot{A} = \left(u\cdot \partial\right) A$ for the co-moving derivatives, $\nabla^\mu \equiv \Delta^{\mu\alpha} \partial_\alpha$ for spacelike derivatives, $\theta \equiv \partial_\mu u^\mu$ is the expansion scalar and, $\sigma^{\mu\nu} \equiv \Delta^{\mu\nu}_{\alpha\beta} \left(\partial^\alpha u^\beta\right)$ is the velocity stress tensor with $\Delta^{\mu\nu}_{\alpha\beta} \equiv \frac{1}{2} \left(\Delta^\mu_\alpha \Delta^\nu_\beta + \Delta^\mu_\beta \Delta^\nu_\alpha\right) - \frac{1}{3} \Delta^{\mu\nu} \Delta_{\alpha\beta}$ being the doubly symmetric, traceless tensor. 

The equilibrium variables defined above are defined using the equilibrium distribution, which for the present work has been chosen to have the form of Maxwell-Boltzmann as
\begin{align}
    f_{0,s} &= f_{0, \textbf{p}} e^{\frac{1}{2} \left(s:\omega\right)}
    \approx f_{0,\textbf{p}} \left[1 + \frac{1}{2} \left(s:\omega\right)\right] + \mathcal{O}\left(\omega^2\right)
\end{align}
to describe distribution of the particles in the extended phase-space. Here, $f_{0,\textbf{p}} = e^{- \beta \left(u\cdot p\right) + \xi}$ is the usual phase-space distribution function with $\beta = 1/T$ being the inverse temperature and, $\xi = \mu/T$ being the ratio of chemical potential to temperature. Note that, in the last step, we have used the small polarization limit keeping terms only linear in $\omega^{\mu\nu}$, which is the Lagrange multiplier associated with the consideration of angular momentum \cite{Becattini:2018duy}. To obtain the distribution for the anti-particles we have to substitute $\xi \to - \xi$. Then we may obtain the equilibrium variables as,
\begin{subequations}
    \begin{align}
        n_0 &= u_\mu N^{\mu}_0 = \ab{\left(u\cdot p\right) \left(1 + \frac{1}{2} s_{\alpha\beta} \omega^{\alpha\beta}\right)}_0^-
        = 2 I_{10}^-, \label{n0-def}\\
        \varepsilon_0 &= u_\mu u_\nu T^{\mu\nu}_0 = \ab{\left(u\cdot p\right)^2 \left(1 + \frac{1}{2} s_{\alpha\beta} \omega^{\alpha\beta}\right)}_0^+
        = 2 I_{20}^+, \label{ve0-def}\\
        P_0 &= - \frac{\Delta_{\mu\nu}}{3} T^{\mu\nu}_0 = - \frac{1}{3} \ab{\left(p \cdot \Delta \cdot p\right) \left(1 + \frac{1}{2} s_{\alpha\beta} \omega^{\alpha\beta}\right)}_0^+
        = - 2\, I_{21}^+, \label{P0-def}\\
        S_0^{\lambda,\mu\nu} &= \ab{p^\lambda s^{\mu\nu} \left(1 + \frac{1}{2} s_{\alpha\beta} \omega^{\alpha\beta}\right)}_0^+
        =  \frac{1}{m^2} \left[ G_{I,10}^+ u^\lambda \omega^{\mu\nu} + 2 H_{I,30}^+ u^{\lambda} u^\alpha u^{[\mu}\right. 
        + \left.2 I_{31}^+ \left(u_\alpha \Delta^{\lambda[\mu} + \Delta^{\lambda}_{\alpha} u^{[\mu}\right) \omega^{\nu]\alpha}\right], \label{S_0-def}
    \end{align}
\end{subequations}
where we have used the notations,
\begin{subequations}
    \begin{align}
        \ab{\left(\cdots\right)}_0^\pm &= \ab{\left(\cdots\right)}_0 \pm \ab{\left(\cdots\right)}_{\Bar{0}}, \label{bracket1-def}\\
        \ab{\left(\cdots\right)}_0 &= \int dP dS \left(\cdots\right) f_{0,\textbf{p}}, \label{bracket2-def}\\
        \ab{\left(\cdots\right)}_{\Bar{0}} &= \int dP dS \left(\cdots\right) \Bar{f}_{0,\textbf{p}}, \label{bracket3-def}
    \end{align}
\end{subequations}
along with the thermodynamic integrals which are defined as,
\begin{subequations}
    \begin{align}
        Y_{(r)\pm}^{\mu_1 \mu_2 \cdots \mu_n} (X) &= \int dP\, X \left(u\cdot p\right)^{r} p^{\mu_1} p^{\mu_2} \cdots p^{\mu_n} \left(f_{0, \textbf{p}} \pm \Bar{f}_{0, \textbf{p}}\right), \label{Y-int-def} \\
        Y_{nj}^\pm(X) &= \frac{1}{\left(2j+1\right)!!} \int dP\, X \left(u\cdot p\right)^{n-2j} \left(p\cdot \Delta \cdot p\right)^j \left(f_{0, \textbf{p}} \pm \Bar{f}_{0, \textbf{p}}\right), \label{TI-def1}
    \end{align}
\end{subequations}
for some scalar function $X$. This allows us to identify, $I_{nj}^\pm = Y_{nj}^\pm(1)$. Similar relations hold for the tensorial integrals in Eq.~\eqref{Y-int-def}. Also note that, the $G$ and $H$-type thermodynamic integrals are defined in Eq.~\eqref{G,H-def}.

We can further define the out-of-equilibrium currents using the fact that the out-of-equilibrium distribution function of a system close to equilibrium can be written as, $f = f_0 + \delta f = f_0 \left(1 + \phi\right)$ where $|{\delta f}| \ll f_0$. This holds for anti-particles as well with $f_0 \to \Bar{f}_0$, $\delta f \to \delta\Bar{f}$ and, $\phi \to \Bar{\phi}$. Then we can write,
\begin{subequations}
    \begin{align}
        &\delta n = u_\mu \delta N^\mu = \ab{\left(u\cdot p\right) \phi_s}_0 - \ab{\left(u\cdot p\right) \Bar{\phi}_s}_{\Bar{0}}, \label{dn}\\
        &\delta \varepsilon = u_\mu u_\nu \delta T^{\mu\nu} = \ab{\left(u\cdot p\right)^2 \phi_s}_0 + \ab{\left(u\cdot p\right)^2 \Bar{\phi}_s}_{\Bar{0}}, \label{de}\\
        &\delta P = - \frac{\Delta_{\mu\nu}}{3} \delta T^{\mu\nu}
        = - \frac{1}{3} \Big[\ab{\left(p \cdot \Delta \cdot p\right) \phi_s}_0 + \ab{\left(p \cdot \Delta \cdot p\right) \Bar{\phi}_s}_{\Bar{0}} \Big], \label{dP}\\
        &n^\mu = \Delta^{\mu}_{\nu} \delta N^{\nu} = \ab{p^{\ab{\mu}} \phi_s}_0 - \ab{p^{\ab{\mu}} \Bar{\phi}_s}_{\Bar{0}} \label{n^m}\\
        &h^\alpha = u_\mu \Delta^{\alpha}_{\nu} \delta T^{\mu\nu}
        = \ab{\left(u \cdot p\right) p^{\ab{\alpha}} \phi_s}_0 + \ab{\left(u \cdot p\right) p^{\ab{\alpha}} \Bar{\phi}_s}_{\Bar{0}}, \label{h^m}\\
        &\pi^{\mu\nu} = \Delta^{\mu\nu}_{\alpha\beta} \delta T^{\alpha\beta} = \ab{p^{\langle\mu} p^{\nu\rangle} \phi_s}_0 + \ab{p^{\langle\mu} p^{\nu\rangle} \Bar{\phi}_s}_{\Bar{0}}, \label{pi^mn}\\
        &\delta S^{\lambda,\mu\nu} = \ab{p^\lambda s^{\mu\nu} \phi_s}_0 + \ab{p^\lambda s^{\mu\nu} \Bar{\phi}_s}_{\Bar{0}}, \label{S^lmn}
    \end{align}
\end{subequations}
where we have used the notations, $A^{\ab{\alpha}} = \Delta^\alpha_\mu A^\mu$ and $A^{\langle\mu} B^{\nu\rangle} = \Delta^{\mu\nu}_{\alpha\beta} A^\alpha B^\alpha$. To obtain the explicit expressions of these currents, we need the specific forms of $\phi_s$ and $\Bar{\phi}_s$. These can be found by solving the relativistic Boltzmann equation. In the next section, we will do that by using the extended relaxation time approximation.

\section{Transport Properties}
\label{sec:TP}

\subsection{Boltzmann Equation}

The Boltzmann equation with extended relaxation time approximation is given by \cite{Dash:2021ibx},
\begin{align}
    p^\mu \partial_\mu f_{s} &= - \frac{\left(u\cdot p\right)}{\tau_{\rm R}(x,p,s)} \left(f_{s} - f_{0,s}^*\right), \label{Beq1}
\end{align}
where $\tau_{\rm R}(x,p,s)$ is the relaxation time that may depend on spacetime, momentum as well as the spin of the particle and, $f_{0,s}^*$ is the extended phase-space distribution function in thermodynamic equilibrium defined as (for discussions on the difference between the thermodynamic and hydrodynamic equilibrium, see Ref.~\cite{Dash:2021ibx}),
\begin{align}
    f_{0,s}^* =f_{0,\textbf{p}}^* \left[1 + \frac{1}{2} s^{\alpha\beta} \omega_{\alpha\beta}^*\right] + \mathcal{O}\left(\omega^2\right),
\end{align}
where $f_{0,\textbf{p}}^* = e^{- \beta^*\left(u^*\cdot p\right) + \xi^*}$ with $\beta^* = 1/T^*$, $\xi^* = \mu^*/T^*$ being the inverse temperature and ratio of chemical potential to temperature in the thermodynamic equilibrium whereas $u^*_\mu$ and $\omega^*_{\alpha\beta}$ are the thermodynamic fluid four-velocity and spin polarization tensor respectively. The hydrodynamic (unstarred) and thermodynamic (starred) variables are related to each other as,
\begin{subequations}
    \begin{align}
        T^* &\equiv T + \delta T,
        ~~\qquad
        \mu^* \equiv \mu + \delta \mu, \label{T*,mu*}\\
        u_\mu^* &\equiv u_\mu + \delta u_\mu,
        ~~\quad
        \omega_{\mu\nu}^* \equiv \omega_{\mu\nu} + \delta \omega_{\mu\nu}. \label{u*,om*}
    \end{align}
\end{subequations}
The quantities $\delta T, \delta \mu, \delta u_\mu$ and $\delta\omega_{\mu\nu}$ are at lease first order in spacetime gradients.

To solve the Boltzmann equation, we adopt the Chapman-Enskog like iterative expansion up to first order in spacetime gradient. Using $f_s = f_{0,s} + \delta f_{1,s}$ and, $f_{0,s}^* = f_{0,s} + \delta f_{1,s}^*$ we can re-write the Boltzmann equation as,
\begin{align}
    p^\mu \partial_\mu f_{0,s} &= - \frac{\left(u\cdot p\right)}{\tau_{\rm R}(x,p,s)} \left(\delta f_{1,s} - \delta f_{1,s}^*\right). \label{Beq2}
\end{align}
We can obtain the expression of $\delta f_{1,s}^*$ by performing Taylor expansion of $f_{0,s}^*$ around $T, \mu, u_\mu$ and $\omega_{\mu\nu}$ as,
\begin{align}
    \delta f_{1,s}^* &= \left[ \left\{- \frac{p^\mu \delta u_\mu}{T} + \left(\frac{\left(u\cdot p\right) - \mu}{T^2}\right) \delta T + \frac{\delta \mu}{T} \right\} \left(1 + \frac{1}{2} \omega_{\mu\nu} s^{\mu\nu}\right) + \frac{1}{2} s^{\mu\nu} \delta \omega_{\mu\nu} \right] f_{0,\textbf{p}}. \label{df*1}
\end{align}
The same for anti-particles can be obtained from Eq.~\eqref{df*1} by substituting $\mu \to - \mu$, $\delta\mu \to - \delta\mu$ and $f_{0,\textbf{p}} \to \Bar{f}_{0,\textbf{p}}$.

The quantities, $\delta T, \delta \mu, \delta u_\mu$ and $\delta\omega_{\mu\nu}$ can be determined from the frame and matching conditions which we express as,
\begin{subequations}
    \begin{align}
        \ab{q_1\, \phi_s^+}_0 + \ab{\Bar{q}_1\, \phi_s^-}_{\Bar{0}} &= 0, \label{mc1}\\
        \ab{q_2\, \phi_s^+}_0 + \ab{\Bar{q}_2\, \phi_s^-}_{\Bar{0}} &= 0, \label{mc2}\\
        \ab{q_3\, p^{\ab{\mu}} \phi_s^+}_0 + \ab{\Bar{q}_3\, p^{\ab{\mu}} \phi_s^-}_{\Bar{0}} &= 0, \label{mc3}\\
        \ab{q_4\, s^{\mu\nu} \phi_s^+}_0 + \ab{\Bar{q}_4\, s^{\mu\nu} \phi_s^-}_{\Bar{0}} &= 0, \label{mc4}
    \end{align}
\end{subequations}
where the quantities, $q_r$ and $\Bar{q}_r$ for $r=1,2,3,4$ are functions of spacetime and momentum\footnote{While we can also consider $q_r$ and $\Bar{q}_r$ to depend on spin as well, in the present case we do not do this for the sake of simplicity.}. For Landau-Lifshitz (LL) frame \cite{landau1987fluid} we have to set, $q_1 = - \Bar{q}_1 = \left(u\cdot p\right)$, $q_2 = \Bar{q}_2 = \left(u\cdot p\right)^2$, $q_3 = \Bar{q}_3 = \left(u\cdot p\right)$ and, $q_4 = \Bar{q}_4 = \left(u\cdot p\right)$. Using Eqs.~\eqref{Beq2} and \eqref{df*1} in Eqs.~\eqref{mc1}-\eqref{mc4} we find,
\begin{subequations}
    \begin{align}
        \delta u^\mu &= \beta\, \mathcal{C}_1 \left(\nabla^\mu\xi\right),
        \qquad
        \delta T = \mathcal{C}_2\, \theta,
        \qquad
        \delta \mu = \mathcal{C}_3\, \theta, \label{dT,m,u-expression}\\
        \delta \omega^{\mu\nu} &= \mathcal{D}_\Pi^{\mu\nu} \theta + \mathcal{D}_n^{\mu\nu\gamma} \left(\nabla_\gamma \xi\right) + \mathcal{D}_\pi^{\mu\nu\alpha\beta} \sigma_{\alpha\beta} + \mathcal{D}_\Sigma^{\mu\nu\gamma\alpha\beta} \left(\nabla_\gamma \omega_{\alpha\beta}\right), \label{domega-expression}
    \end{align}
\end{subequations}
The detailed expressions of the coefficients are given Appendix \ref{app1}, which were derived by assuming $q_1 = - \Bar{q}_1$, and $q_{2/3/4} = \Bar{q}_{2/3/4}$. In the process of obtaining Eqs.~\eqref{dT,m,u-expression} and \eqref{domega-expression}, we used the following relations,
\begin{subequations}
    \begin{align}
        \Dot{\beta} \!&=\! \chi_b\, \theta,
        \qquad
        \Dot{\xi} \!=\! \chi_a\, \theta,
        \qquad
        \beta \Dot{u}_\mu \!=\! - \left(\nabla_\mu \beta\right) +\! \frac{\left(\nabla_\mu \xi\right)}{h_0}, \label{D-b,T,m}\\
        \Dot{\omega}^{\mu\nu} &= D_\Pi^{\mu\nu}\, \theta + D_n^{\mu\nu\alpha} \left(\nabla_\alpha\xi\right) + D_\pi^{\mu\nu\alpha\beta}\, \sigma_{\alpha\beta} + D_\Sigma^{\mu\nu\alpha\beta\gamma} \left(\nabla_\alpha \omega_{\beta\gamma}\right) \label{D-omega}
    \end{align}
\end{subequations}
which were obtained from the hydrodynamic equations \eqref{Heq1}-\eqref{Heq4}. The coefficients $\chi_a$ and $\chi_b$ are given in Appendix \ref{app1} whereas the $D$-type coefficients of Eq.~\eqref{D-omega} are the same as the ones found in Ref.~\cite{Bhadury:2020cop}. Once we have the expression for $\delta f^*_{1,s}$, we can use Eq.~\eqref{Beq2} to solve for $\delta f_{1,s}$ as,
\begin{align}
    \delta f_{1,s} = \left[\mathcal{A}_{s} + \mathcal{B}_{s} + \mathcal{C}_{s} + s^{\alpha\beta} \mathcal{D}_{s,\,\alpha\beta}\right]\! f_{0,\textbf{p}}, \label{df1s}
\end{align}
where,
\begin{subequations}
    \begin{align}
        \mathcal{A}_{s} &= - \tau_{\rm R} \left[ \left(\frac{\beta}{3} - \chi_b\right) \left(u\cdot p\right) - \frac{\beta\, m^2}{3 \left(u\cdot p\right)} + \chi_a \right] \left(1 + \frac{1}{2} \omega_{\alpha\beta} s^{\alpha\beta}\right) \theta - \frac{\tau_{\rm R} }{2} s_{\alpha\beta}\, D_{\Pi}^{\alpha\beta}\, \theta + a_s, \label{til-A_s}\\
        \mathcal{B}_{s} &= \tau_{\rm R} \!\left( h_0^{-1} - \frac{1}{\left(u\cdot p\right)} \right)\!\! \left(1 + \frac{1}{2} \omega_{\alpha\beta} s^{\alpha\beta}\right)\! p^{\ab{\mu}} \!\left(\nabla_\mu \xi\right) - \frac{\tau_{\rm R} }{2} s_{\alpha\beta}\, D_{n}^{\alpha\beta\mu} \left(\nabla_\mu\xi\right) + p^{\ab{\mu}} b_{s,\,\mu}, \label{til-B_s}\\
        \mathcal{C}_{s} &= \frac{\beta \tau_{\rm R}}{\left(u\cdot p\right)} \left(1 + \frac{1}{2} s_{\alpha\beta} \omega^{\alpha\beta}\right) p^{\langle\mu} p^{\nu\rangle} \sigma_{\mu\nu} - \frac{\tau_{\rm R}}{2} s_{\alpha\beta} D_{\pi}^{\alpha\beta\mu\nu}\, \sigma_{\mu\nu}, \label{til-C_s}\\
        \mathcal{D}_{s,\,\alpha\beta} &= \frac{1}{2} \bigg[ - \frac{\tau_{\rm R} p^{\ab{\mu}}}{\left(u\cdot p\right)} \left(\nabla_\mu\omega_{\alpha\beta}\right) - \tau_{\rm R} D_{\Sigma,\,\alpha\beta\gamma\mu\nu} \left(\nabla^\gamma\omega^{\mu\nu}\right) + 2 c_{s,\,\alpha\beta} \bigg]. \label{til-D_s}
    \end{align}
\end{subequations}
and we have used the following abbreviations,
\begin{subequations}
    \begin{align}
        a_s &=\! \left[\! \left\{\!\frac{\left(u\cdot p\right) - \mu}{T^2} \!\right\} \delta T \!+\! \frac{\delta \mu}{T} \right]\! \!\left(\!1 + \frac{1}{2} \omega_{\mu\nu} s^{\mu\nu}\!\right), \label{a_s}\\
        b_{s,\,\mu} &= - \beta\, \delta u_\mu \left(1 + \frac{1}{2} \omega_{\alpha\beta} s^{\alpha\beta} \right), \label{b_s}\\
        c_{s,\,\mu\nu} &= \delta \omega_{\mu\nu}/2. \label{c_s}
    \end{align}
\end{subequations}
The expressions for anti-particle versions of Eqs.~\eqref{df1s}-\eqref{c_s} can be obtained thourgh the following substitutions, $f_{0,\textbf{p}} \to \Bar{f}_{0,\textbf{p}}$ in Eq.~\eqref{df1s}, $\chi_a \to - \chi_a$ in Eq.~\eqref{til-A_s}, $- \left(u\cdot p\right)^{-1} \to \left(u\cdot p\right)^{-1}$ in Eq.~\eqref{til-B_s}, $\mu \to - \mu$ and $\delta \mu \to - \delta \mu$ in Eq.~\eqref{a_s}. 

\subsection{Transport Coefficients}
\label{ssec:TC}

In the last part, we obtained the non-equilibrium corrections to the distribution functions and now we can use it to obtain the transport coefficients. Then the relativistic Navier-Stokes equations for the non-equilibrium currents are given by,
\begin{subequations}
    \begin{align}
        \delta n &= \nu\, \theta,
        \quad
        \delta \varepsilon = e\, \theta,
        \quad
        \delta P = \rho\, \theta, \label{del_n,e,P-NS}\\
        n^\mu &= \kappa_n^{\mu\nu} \left(\nabla_\nu\xi\right),
        \quad
        h^\mu = \kappa_h^{\mu\nu} \left(\nabla_\nu\xi\right), \label{n^m,h^m-NS}\\
        \pi^{\mu\nu} &= 2 \eta^{\mu\nu\alpha\beta} \sigma_{\alpha\beta}, \label{pi^mn-NS}\\
        \delta S^{\lambda,\mu\nu} &= B_\Pi^{\lambda\mu\nu} \theta + B_n^{\lambda\mu\nu\alpha} \left(\nabla_\alpha\xi\right) + B_\pi^{\lambda\mu\nu\alpha\beta} \sigma_{\alpha\beta} + B_\Sigma^{\lambda\mu\nu\gamma\alpha\beta} \left(\nabla_\gamma\omega_{\alpha\beta}\right), \label{del_S^lmn-NS}
    \end{align}
\end{subequations}
where the transport coefficients are given in Appendix \ref{app1}. The crucial thing to note here is that not all these transport coefficients are invariant under a change of frame and matching conditions. We can determine the frame and matching invariant transport coefficients from calculating the entropy production as described in the next part. Another important issue is to note that there is no influence of spin on the evolution of the fluid in this approach, where the evolution of spin does depend on the dynamical evolution of the fluid background. One may, however, find the spin to affect the fluid evolution if either the effect of an electromagnetic field is considered \cite{Bhadury:2022ulr} or if the effect of quadratic polarization (i.e. keeping $\omega^2$ terms and neglecting $\mathcal{O}(\omega^3)$ terms) is taken into consideration \cite{Florkowski:2024bfw, Drogosz:2024gzv}.

Note that so far we have not specified the functional dependency of the relaxation time. In Section.~\ref{ssec:SC}, we will assume the following power-law parametrization scheme for the relaxation time as,
\begin{align}
    \tau_{\rm R} (x,p,s) = \tau_{\rm eq} (x,p) \left(u\cdot s\right)^{2 \ell}. \label{tau-parameter}
\end{align}
where $\tau_{\rm eq}(x,p)$ is independent of spin and $\ell (\in \mathbb{Z}^+)$ is assumed to be positive integer. The reason of this choice is that with fractional power of $\left(u\cdot p\right)$ the integrals cannot be performed analytically and the factor of $2$ is chosen as the spin integrals with an odd number of $s^\mu$ vanish.

\subsection{Entropy Production}
\label{ssec:EP}

As mentioned before, the transport coefficients calculated in the last section, are not frame-invariant. The frame-invariant transport coefficients can be determined by evaluating the entropy production. The entropy current can be found to be given by,
\begin{align}
    \mathcal{H}^\mu = \beta_\nu T^{\mu\nu} - \xi N^\mu - \frac{1}{2} \omega_{\alpha\beta} S^{\mu,\alpha\beta} + P_0 \beta^\mu. \label{H^mu-def}
\end{align}
Then, we can calculate the divergence of the entropy current, $\partial_\mu \mathcal{H}^\mu$ by noting \cite{Florkowski:2024bfw},
\begin{align}
    \partial_\mu \!\left(\!P_0 \beta^\mu\right) \!=\! N_0^\mu \!\left(\partial_\mu \xi\right) \!-\! T_0^{\mu\lambda} \!\left(\!\partial_\mu \beta_\lambda\right) \!+\! \frac{1}{2} S_0^{\mu,\alpha\beta} \!\left(\!\partial_\mu \omega_{\alpha\beta}\right)\!, \label{d(Pb)}
\end{align}
as,
\begin{align}
    \partial_\mu \mathcal{H}^\mu &=\! \left(\partial_\mu\beta_\nu\right)\! T^{\mu\nu} \!-\! \left(\partial_\mu \xi\right)\! N^\mu \!-\! \frac{1}{2} \left(\partial_\mu \omega_{\alpha\beta}\right)\! S^{\mu,\alpha\beta} \!+\! \partial_\mu \!\left(P_0 \beta^\mu\right) \nonumber\\
    &= \left(\chi_b\, \delta \varepsilon \!-\! \beta \delta P \!-\! \chi_a\, \delta n\right)\! \theta \!+\! \frac{1}{h_0} \!\left(h^{\mu} - h_0 n^\mu\right)\! \left(\nabla_\mu \xi\right) \!+\! \beta \pi^{\mu\nu} \sigma_{\mu\nu} - \frac{1}{2} \left(u^\nu D_\Sigma^{\rho\lambda\mu\alpha\beta} \delta S_{\nu,\rho\lambda} + \delta S^{\mu,\alpha\beta}\right) \left(\nabla_\mu \omega_{\alpha\beta}\right), \label{dH1}
\end{align}
Defining,
\begin{align}
    \partial_\mu \mathcal{H}^\mu \!=\! \beta \pi^{\mu\nu} \!\sigma_{\mu\nu} \!-\! \beta \Pi \theta \!-\! \mathcal{Q}^\mu \!\left(\nabla_{\!\mu} \xi\right) \!-\! \mathcal{S}^{\mu\alpha\beta} \!\left(\nabla_{\!\mu} \omega_{\alpha\beta}\right)\!, \label{dH2}
\end{align}
we can write frame-invariant dissipative currents as,
\begin{subequations}
    \begin{align}
    &\Pi = - \zeta \theta = \delta P - \frac{\chi_b}{\beta} \delta \varepsilon + \frac{\chi_a}{\beta} \delta n
    = \left(\rho - \frac{\chi_b\, e}{\beta} + \frac{\chi_a\, \nu}{\beta}\right) \theta, \label{Pi1}\\
    &\mathcal{Q}^\mu = n^\mu - \frac{h^\mu}{h_0} = \left(\kappa_n^{\mu\nu} - \frac{1}{h_0} \kappa_h^{\mu\nu}\right) \left(\nabla_\nu \xi\right)
    = \kappa^{\mu\nu} \left(\nabla_\nu\xi\right), \label{Q1}\\
    &\pi^{\mu\nu} = 2 \eta^{\mu\nu\alpha\beta} \sigma_{\alpha\beta}, \label{pi1}\\
    &\mathcal{S}^{\mu\alpha\beta} = \frac{1}{2} \left(u^\nu D_\Sigma^{\rho\lambda\mu\alpha\beta} \delta S_{\nu,\rho\lambda} + \delta S^{\mu,\alpha\beta}\right)
    = \frac{1}{2} \!\left(u^\nu D_\Sigma^{\rho\lambda\mu\alpha\beta} B_{\Sigma,\,\nu\rho\lambda}^{~~~~~~\gamma\phi\varphi} + B_\Sigma^{\mu\alpha\beta\gamma\phi\varphi}\right)\! \left(\nabla_\gamma\omega_{\phi\varphi}\right)
    = \beta_\Sigma^{\mu\alpha\beta\gamma\phi\varphi} \left(\nabla_\gamma\omega_{\phi\varphi}\right). \label{S1}
\end{align}
\end{subequations}
Then we can write,
\begin{subequations}
    \begin{align}
        \zeta &= \frac{\chi_b\, e}{\beta} - \rho - \frac{\chi_a\, \nu}{\beta}, \label{zeta1}\\
        \kappa^{\mu\nu} &= \kappa_n^{\mu\nu} - \kappa_h^{\mu\nu}/h_0, \label{kappa1}\\
        \beta_\Sigma^{\mu\alpha\beta\gamma\phi\varphi} &= \frac{1}{2} \left(u^\nu D_\Sigma^{\rho\lambda\mu\alpha\beta} B_{\Sigma,\,\nu\rho\lambda}^{~~~~~~\gamma\phi\varphi} + B_\Sigma^{\mu\alpha\beta\gamma\phi\varphi}\right). \label{beta_S1}
    \end{align}
\end{subequations}
Eq.~\eqref{beta_S1} is the frame-invariant spin transport coefficient and shows to be the only dissipative transport coefficient associated with spin. It is straightforward to note from the first equality of Eq.~\eqref{S1} that the simple choice of $q_4 = \left(u\cdot p\right)$, which implies $u_\nu \delta S^{\nu,\rho\lambda} = 0$, along with other LL frame choices gives back the results of Ref.~\cite{Bhadury:2020cop}.

\subsection{Special Cases}
\label{ssec:SC}

Next, we examine the cases for some specific values of the spin dependence of the relaxation time. We will study a few cases in the following for different values of $\ell$. Furthermore, in the following, we will be using the usual LL frame and matching conditions, that are mentioned in the paragraph between Eqs.~\eqref{mc4} and \eqref{dT,m,u-expression} to make the comparison with previous works easy, however, within the present framework any other matching conditions could be chosen.

\begin{widetext}
    
\subsubsection{\texorpdfstring{$\ell = 0$}{~} case :}
\label{ell=0}

This case can be thought of as the simple inclusion of spin in the framework of ERTA, where the relaxation depends only on spacetime and momenta of constituent particles. In this case the coefficients $\mathcal{C}_{1,2,3}$ turn back to the ones we obtained in Ref.~\cite{Dash:2021ibx} and The $\mathcal{D}$-type coefficients from Eq.~\eqref{domega-expression} reduces to:
\begin{subequations}
    \begin{align}
        \mathcal{D}_\Pi^{\mu\nu} &= \frac{2}{G_{I,10}^+} \bigg[ H_{I,30}^+ u^{[\mu} \mathcal{C}_\Pi^{\nu]} + G_{K,10}^+ D_{\Pi}^{\mu\nu} + 2 H_{K,30}^+ u_\alpha u^{[\mu} D_\Pi^{\nu]\alpha} \nonumber\\
        &- \bigg\{ \left(\chi_b - \frac{\beta}{3}\right) G_{K,20}^+ + \frac{\beta\, m^2}{3} G_{K,00}^+ - \chi_a G_{K,20}^- + \left(G_{I,20}^+ - \mu G_{I,10}^- \right) \frac{\mathcal{C}_2}{T^2} + G_{I,10}^- \frac{\mathcal{C}_3}{T} \bigg\} \omega^{\mu\nu} \nonumber\\
        &- \bigg\{ \left(\chi_b - \frac{\beta}{3}\right) H_{K,40}^+ + \frac{\beta\, m^2}{3} H_{K,20}^+ - \chi_a H_{K,40}^- + \left(H_{I,40}^+ - \mu H_{I,30}^- \right) \frac{\mathcal{C}_2}{T^2} + H_{I,30}^- \frac{\mathcal{C}_3}{T} \bigg\} u_\alpha u^{[\mu}\omega^{\nu]\alpha} \bigg], \label{D_Pi-l0}\\
        \mathcal{D}_n^{\mu\nu\gamma} &= \frac{2}{G_{I,10}^+} \bigg[ H_{I,30}^+ u^{[\mu} \mathcal{C}_{n}^{\nu]\gamma} + G_{K,10}^+ D_{n}^{\mu\nu\gamma} + 2 H_{K,30}^+ u_\alpha u^{[\mu} D_n^{\nu]\alpha\gamma} \nonumber\\
        &- \frac{1}{2} \left(\frac{K_{41}^+}{h_0} - K_{31}^- - \beta^2\, \mathcal{C}_1 I_{41}^+\right) \epsilon^{\mu\nu\rho\phi} \epsilon^{\alpha\beta\lambda}_{~~~~~\phi} \left(u_\rho \Delta^\gamma_\lambda + u_\lambda \Delta^\gamma_\rho\right) \omega_{\alpha\beta} \bigg], \label{D_n-l0}\\
        \mathcal{D}_\pi^{\mu\nu\alpha\beta} &= \frac{2}{G_{I,10}^+} \bigg[ H_{I,30}^+ u^{[\mu} \mathcal{C}_\pi^{\nu]\alpha\beta} + G_{K,10}^+ D_{\pi}^{\mu\nu\alpha\beta} + 2 H_{K,30}^+ u_\gamma u^{[\mu} D_\pi^{\nu]\gamma\alpha\beta} - 4 \beta\, K_{42}^+ \Delta^{\alpha\beta,\gamma[\mu} \omega^{\nu]}_{~~\gamma} \bigg], \label{D_pi-l0}\\
        \mathcal{D}_\Sigma^{\mu\nu\gamma\alpha\beta} &= \frac{2}{G_{I,10}^+} \bigg[ H_{I,30}^+ u^{[\mu} \mathcal{C}_\Sigma^{\nu]\gamma\alpha\beta} + G_{K,10}^+ D_{\Sigma}^{\mu\nu\gamma\alpha\beta} + 2 H_{K,30}^+ u_\rho u^{[\mu} D_\Sigma^{\nu]\rho\gamma\alpha\beta} - \frac{K_{31}^+}{2} \epsilon^{\mu\nu\rho\phi} \epsilon^{\alpha\beta\lambda}_{~~~~~\phi} \left(u_\rho \Delta^\gamma_\lambda + u_\lambda \Delta^\gamma_\rho\right) \bigg], \label{D_Sig-l0}
    \end{align}
\end{subequations}
where we have used the notation, $\Delta^{\mu\nu,\alpha\beta} = \Delta^{\mu\nu}_{\gamma\phi} g^{\gamma\alpha} g^{\phi\beta}$. The $\mathcal{C}$-type coefficients defined above can be simplified from Eqs.~\eqref{C_Pi-gen}-\eqref{C_Sig-gen} to,
\begin{subequations}
    \begin{align}
        \mathcal{C}_\Pi^\nu &= - \frac{1}{I_{31}^+} \bigg[\!\! \left\{\!\! \left(\chi_b \!-\! \frac{\beta}{3}\right) G_{K,20}^+ + \frac{\beta m^2}{3} G_{K,00}^+ - \chi_a G_{K,20}^- + \beta^2 \left( G_{I,20}^+ - \mu G_{I,10}^-\right) \mathcal{C}_2 + \beta G_{I,10}^- \mathcal{C}_3\right\} u_\mu \omega^{\mu\nu} \nonumber\\
        &+ 2 \left\{ \left(\chi_b \!-\! \frac{\beta}{3}\right) H_{K,40}^+ + \frac{\beta m^2}{3} H_{K,20}^+ - \chi_a H_{K,40}^- + \beta^2 \left( H_{I,40}^+ - \mu H_{I,30}^-\right) \mathcal{C}_2 + \beta H_{I,30}^- \mathcal{C}_3 \right\} u_\mu u_\alpha u^{[\mu} \omega^{\nu]\alpha} \nonumber\\
        &- G_{K,10}^+ u_\mu D_\Pi^{\mu\nu} - 2 H_{K,30}^+ u_\mu u_\alpha u^{[\mu} D_\Pi^{\nu]\alpha}, \label{C_Pi-l0}\\
        \mathcal{C}_n^{\nu\gamma} &= - \frac{2}{I_{31}^+} \bigg[\! \left( \frac{K_{41}^+}{h_0} - K_{31}^- - \beta^2\, \mathcal{C}_1 I_{41}^+ \right) u_\mu \Delta^\gamma_\alpha u^{[\mu} \omega^{\nu]\alpha} - G_{K,10}^+ u_\mu D_n^{\mu\nu\gamma} - 2 G_{K,30}^+ u_\mu u_\alpha u^{[\mu} D_n^{\nu]\alpha\gamma} \!\bigg], \label{C_n-l0}\\
        \mathcal{C}_\pi^{\nu\gamma\lambda} &= \frac{2}{I_{31}^+} \bigg[ \beta\, K_{42}^+ \Delta^{\gamma\lambda,\alpha\nu} u_\mu \omega^{\mu}_{~~\alpha} + K_{41}^+ u_\mu \left( u_\alpha \Delta^{\gamma[\mu} + \Delta^\gamma_\alpha u^{[\mu}\right) D_\pi^{\nu]\alpha\gamma\lambda} \bigg], \label{C_pi-l0}\\
        \mathcal{C}_\Sigma^{\nu\gamma\phi\varphi} &= \frac{1}{I_{31}^+} \left[ \frac{K_{31}^+}{2} \epsilon^{\mu\nu\gamma\alpha} \epsilon^{\phi\varphi\lambda}_{~~~~~\alpha} u_\mu u_\lambda - G_{K,10}^+ u_\mu D_\Sigma^{\mu\nu\gamma\phi\varphi} - 2 H_{K,30}^+ u_\mu u_\alpha u^{[\mu} D_\Sigma^{\nu]\alpha\gamma\phi\varphi} \right]. \label{C_Sig-l0}
    \end{align}
\end{subequations}
where we made use of the following notations,
\begin{align}
    G_{X,nj}^\pm = m^2 X_{nj}^\pm - 2 X_{n+2,j+1}^\pm,
    \qquad
    H_{X,nj}^\pm = X_{nj}^\pm - X_{n,j+1}^\pm, \label{G,H-def}
\end{align}
$K_{n+r,j}^\pm = Y_{nj}^\pm \left((u\cdot p)^r \tau_{\rm R}\right)$ and in the limit of $\tau_{\rm R}$ being independent of momentum, we can write, $K_{nj}^\pm = \tau_{\rm R} I_{nj}^\pm$. Finally, we write the expressions of the transport coefficients,
\begin{subequations}
    \begin{align}
        \nu &= 2 \left[ \!\left(\chi_b - \frac{\beta}{3}\right) K_{20}^- \!+ \frac{\beta\, m^2}{3} K_{00}^- - \chi_a K_{10}^+ + \frac{\mathcal{C}_2}{T^2} \left(K_{20}^- - \mu K_{10}^+\right) + \frac{\mathcal{C}_3}{T} K_{10}^- \right], \\
        e &= 2 \left[ \!\left(\chi_b - \frac{\beta}{3}\right) K_{30}^+ \!+ \frac{\beta\, m^2}{3} K_{10}^+ - \chi_a K_{20}^- + \frac{\mathcal{C}_2}{T^2} \left(K_{30}^+ - \mu K_{20}^-\right) + \frac{\mathcal{C}_3}{T} K_{20}^+ \right], \\
        \rho &= - 2 \left[ \!\left(\chi_b - \frac{\beta}{3}\right) K_{31}^+ \!+ \frac{\beta\, m^2}{3} K_{11}^+ - \chi_a K_{21}^- + \frac{\mathcal{C}_2}{T^2} \left(K_{31}^+ - \mu K_{21}^-\right) + \frac{\mathcal{C}_3}{T} K_{21}^+ \right], \\
        \kappa_n^{\mu\alpha} &= 2 \left(\frac{1}{h_0} K_{21}^- - \frac{\mathcal{C}_1}{T^2} I_{21}^- - K_{11}^+ \right) \Delta^{\mu\alpha}, \\
        \kappa_h^{\mu\alpha} &= 2 \left(\frac{1}{h_0} K_{31}^+ - \frac{\mathcal{C}_1}{T^2} I_{31}^+ - K_{21}^- \right) \Delta^{\mu\alpha}, \\
        \eta^{\mu\nu\alpha\beta} &= 2 \beta K_{32}^+ \Delta^{\mu\nu,\alpha\beta},
    \end{align}
\end{subequations}
Finally, the spin transport coefficients are given by,
\begin{subequations}
    \begin{align}
        B_\Pi^{\lambda\mu\nu} &= \frac{1}{2} \bigg[ \left\{ \left(\chi_b - \frac{\beta}{3}\right) K^\lambda_{(1)+} + \frac{\beta m^2}{3} K_{(-1)+}^{\lambda} - \chi_a K_{(0)-}^\lambda + \frac{\mathcal{C}_2}{T^2} \left(K_{(0)-}^\lambda - \mu K_{(1)+}^\lambda\right) + \frac{\mathcal{C}_3}{T} K_{(0)+}^\lambda\right\} \omega^{\mu\nu} \nonumber\\
        &\qquad+ \frac{2}{m^2} \left\{ \left(\chi_b - \frac{\beta}{3}\right) K^{\lambda\alpha[\mu}_{(1)+} + \frac{\beta m^2}{3} K_{(-1)+}^{\lambda\alpha[\mu} - \chi_a K_{(0)-}^{\lambda\alpha[\mu} + \frac{\mathcal{C}_2}{T^2} \left(K_{(0)-}^{\lambda\alpha[\mu} - \mu K_{(1)+}^\lambda\right) + \frac{\mathcal{C}_3}{T} K_{(0)+}^{\lambda\alpha[\mu}\right\} \omega^{\nu]}_{~~\alpha} \nonumber\\
        &\qquad- \left(K_{(0)+}^{\lambda} D_{\Pi}^{\mu\nu} - I_{(0)+}^{\lambda} \mathcal{D}_{\Pi}^{\mu\nu}\right) - \frac{2}{m^2} \left( K_{(0)+}^{\lambda\alpha[\mu} D_{\Pi,\alpha}^{\nu]} - I_{(0)+}^{\lambda\alpha[\mu} \mathcal{D}_{\Pi,\alpha}^{\nu]}\right) \bigg], \label{B_Pi-l=0}\\
        B_n^{\lambda\mu\nu\alpha} &= \frac{1}{2} \bigg[ \left(\frac{1}{h_0} K^{\ab{\alpha}\lambda}_{(0)+} - \frac{\mathcal{C}_1}{T^2} I^{\ab{\alpha}\lambda}_{(0)+} - K^{\ab{\alpha}\lambda}_{(-1)-}\right) \omega^{\mu\nu} + \frac{2}{m^2} \left(\frac{1}{h_0} K_{(0)+}^{\ab{\alpha}\lambda\beta[\mu} - \frac{\mathcal{C}_1}{T^2} I_{(0)+}^{\ab{\alpha}\lambda\beta[\mu} -  K_{(-1)-}^{\ab{\alpha}\lambda\beta[\mu}\right) \omega^{\nu]}_{~~\beta} \nonumber\\
        &\qquad- \left(K_{(0)+}^{\lambda} D_n^{\mu\nu\alpha} - K_{(0)+}^{\lambda} \mathcal{D}_n^{\mu\nu\alpha}\right) - \frac{2}{m^2} \left(K_{(0)+}^{\lambda\beta[\mu} D^{\nu]}_{n,\,\beta\phi} g^{\phi\alpha} - I_{(0)+}^{\lambda\beta[\mu} \mathcal{D}^{\nu]}_{n,\,\beta\phi} g^{\phi\alpha}\right) \!\bigg], \label{B_n-l=0}\\
        B_\pi^{\lambda\mu\nu\alpha\beta} &= \frac{\beta}{2} \left[\frac{2}{m^2} \left(K_{(-1)+}^{\ab{\alpha\beta}\lambda\gamma[\mu} \omega^{\nu]}_{~~\gamma} - K_{(0)+,\gamma}^{\lambda[\mu} D_{\pi}^{\nu]\gamma\alpha\beta} + I_{(0)+,\gamma}^{\lambda[\mu} \mathcal{D}_{\pi}^{\nu]\gamma\alpha\beta}\right) - \left(K_{(0)+}^{\lambda} D_\pi^{\mu\nu\alpha\beta} - K_{(0)+}^{\lambda} D_\pi^{\mu\nu\alpha\beta}\right) \right], \label{B_pi-l=0}\\
        B_\Sigma^{\lambda\mu\nu\gamma\alpha\beta} &= \frac{1}{2} \left[ \frac{K_{(-1)+}^{\ab{\gamma}\lambda\rho\phi}}{2 m^2} \epsilon_\rho^{~\,\mu\nu\varphi} \epsilon^{\alpha\beta}_{~~~\phi\varphi} \!-\! \left(\!K_{(0)+}^{\lambda} D_\Sigma^{\mu\nu\gamma\alpha\beta} \!-\! I_{(0)+}^{\lambda} \mathcal{D}_\Sigma^{\mu\nu\gamma\alpha\beta} \!\right) \!- \frac{2}{m^2} \!\left(\! K_{(0)+,\phi}^{\lambda[\mu} D_\Sigma^{\nu]\phi\gamma\alpha\beta} \!- I_{(0)+,\phi}^{\lambda[\mu} \mathcal{D}_\Sigma^{\nu]\phi\gamma\alpha\beta} \right) \!\right]\!. \label{B_Sig-l=0}
    \end{align}
\end{subequations}
Therefore, as expected even for the momentum-dependent relaxation times, there is no feedback of spin on the evolution of fluid although the transport properties are now dependent on the interaction mechanism through the momentum dependence of the relaxation time. Similarly, spin transport also depends on the interaction mechanism, and in addition, as in Ref.~\cite{Bhadury:2020puc, Bhadury:2020cop} they are affected by the evolution of fluid.

\subsubsection{\texorpdfstring{$\ell = 1$}{~} case :}
\label{ell=1}

This case can be easily obtained from the results of Section \ref{ell=0} and Appendix~\ref{app1} by noting that for some spacetime and momentum-dependent function, $X$ we can show the following relation to hold,
\begin{align}
    Y_{nj}^\pm \left(\left(u\cdot s\right)^2 X\right) = - \frac{\left(2j + 1\right)}{2m^2} Y_{n+2,j+1}^\pm \left(X\right). \label{Y_nj-recursion1}
\end{align}
Consequently, we can obtain the results for the $\ell = 1$ case by simply performing the substitution,
\begin{align}
    2m^2\, K_{nj}^\pm \to - \left(2j + 1\right) K_{n+2,j+1}^\pm.
\end{align}
Thus even for this type of spin-dependence of relaxation time, there is no feedback of spin polarization to the evolution of the fluid.

\subsubsection{\texorpdfstring{$\ell = 2$}{~} case :}
\label{ell=2}
Similarly, we can also compute the cases for $\ell = 2$ as well. For this case, we will need to use the relation,
\begin{align}
    \int dS s^\alpha s^\beta s^\mu s^\nu &= \frac{2 \mathfrak{s}^4}{15 m^4} \Big[ 3 p^\alpha p^\beta p^\mu p^\nu \!-\! m^2 \left(p^\alpha p^\beta g^{\mu\nu} + {\rm all~permutations}\right) \!+\! m^4 \left(g^{\alpha\beta} g^{\mu\nu} + g^{\alpha\mu} g^{\beta\nu} + g^{\alpha\nu} g^{\beta\mu}\right)\!\Big], \label{int_s-s^4}
\end{align}
to show that,
\begin{align}
    \int dS \left(u\cdot s\right)^4 = \frac{9}{40 m^4} \left(p \cdot \Delta \cdot p\right)^2. \label{int_s-(u.s)^4}
\end{align}
Then we can prove the relation,
\begin{align}
    Y_{nj}^\pm \left(\left(u\cdot s\right)^4 X\right) = \frac{9\left(2j + 1\right)\left(2j + 3\right)}{40m^4} Y_{n+4,j+2}^\pm \left(X\right). \label{Y_nj-recursion2}
\end{align}
Then through appropriate substitutions, we can similarly obtain the results as in the case of Section \ref{ell=1}. Proceeding in this manner, one may obtain the results for even higher powers.

\end{widetext}

\section{Summary and Outlook}
\label{sec:S&O}

In this article, we have used the ERTA collision kernel to construct a theory of relativistic spin hydrodynamics with momentum and spin-dependent relaxation time. We first examined the case of spin-independent relaxation and then also the case of spin-dependent relaxation time. We found due to the nature of our choice of spin-dependency in Eq.~\eqref{tau-parameter}, that the latter cases can be obtained very easily through a simple substitution of the thermodynamic integrals. We also derived the various transport coefficients for a general frame and matching conditions and expressed the results in terms of thermodynamic integrals through the specific choice of LL frame and matching conditions. We found in the cases of spin-dependence of the relaxation times studied here, the fluid remains unaffected by spin polarization. However, such effects may be observed if we consider a quadratic polarization ($\sim\omega^2$) or include some external field like the electromagnetic fields. While the latter was studied with relaxation time approximation in Ref.~\cite{Bhadury:2022ulr}, the former issue is relatively unexplored \cite{Florkowski:2024bfw, Drogosz:2024gzv} and will be addressed somewhere else. A particularly interesting aspect of the extension to quadratic polarization would be to see whether vorticity can emerge within the kinetic theory with local collisions alone without any consideration of external fields.

The formulation of causal and stable spin hydrodynamics has received little attention and will be investigated in the future. This may include the formulation of a BDNK-type theory for spin hydrodynamics \cite{Weickgenannt:2023btk} or a second-order theory of spin hydrodynamics \cite{Weickgenannt:2022zxs, Biswas:2023qsw}. While there have been developments in this direction, the determination of transport properties requires immediate attention as for numerical implementation of spin hydrodynamics, a causal and stable theory of spin hydrodynamics is of utmost importance.

\section{Acknowledgement}

The author would like to thank Prof. Amaresh Jaiswal and Prof. Wojciech Florkowski for their valuable feedback on the manuscript.

\appendix

\begin{widetext}

\section{List of Various Coefficients}
\label{app1}

In this Appendix, we list various coefficients used throughout the main article. The coefficients defined in Eqs.~\eqref{dT,m,u-expression} and \eqref{domega-expression} are given as,
    \begin{align}
        \mathcal{C}_1 &= \frac{1}{\beta^2\, \ab{q_3 \left(p \cdot \Delta \cdot p\right)/3}_0^+} \left[ \frac{1}{h_0} \ab{ q_3 \tau_{\rm R} \left(p \cdot \Delta \cdot p\right)}_0^+ - \ab{\frac{q_3 \tau_{\rm R}}{\left(u\cdot p\right)} \left(p \cdot \Delta \cdot p\right)}_0^-\right] \\
        \mathcal{C}_2 &= \frac{T^2}{\left[ \ab{q_2 \left(u\cdot p\right)}_0^+ \ab{q_1}_0^+ - \ab{q_1 \left(u\cdot p\right)}_0^- \ab{q_2}_0^- \right]} \left[ \left\{\left(\chi_b - \frac{\beta}{3}\right) \ab{\tau_{\rm R} q_1 \left(u\cdot p\right)}_0^- + \frac{\beta\, m^2}{3} \ab{\frac{\tau_{\rm R} q_1}{\left(u\cdot p\right)}}_0^- - \chi_a \ab{\tau_{\rm R} q_1}_0^+\right\} \ab{q_2}_0^- \right. \nonumber\\
        &\left. - \left\{ \left(\chi_b - \frac{\beta}{3}\right) \ab{\tau_{\rm R} q_2 \left(u\cdot p\right)}_0^+ + \frac{\beta\, m^2}{3} \ab{\frac{\tau_{\rm R} q_2}{\left(u\cdot p\right)}}_0^+ - \chi_a \ab{\tau_{\rm R} q_2}_0^- \right\} \ab{q_1}_0^+ \right] \\
        \mathcal{C}_3 &= \frac{T}{\left[ \ab{q_1 \left(u\cdot p\right)}_0^- \ab{q_2}_0^- - \ab{q_2 \left(u\cdot p\right)}_0^+ \ab{q_1}_0^+ \right]} \nonumber\\
        &\qquad\times\left[\left\{\left(\chi_b - \frac{\beta}{3}\right) \ab{\tau_{\rm R} q_1 \left(u\cdot p\right)}_0^- + \frac{\beta\, m^2}{3} \ab{\frac{\tau_{\rm R} q_1}{\left(u\cdot p\right)}}_0^- - \chi_a \ab{\tau_{\rm R} q_1}_0^+\right\} \left\{ \ab{q_2 \left(u\cdot p\right)}_0^+ - \mu \ab{q_2}_0^- \right\} \right. \nonumber\\
        &\qquad- \left. \left\{\left(\chi_b - \frac{\beta}{3}\right) \ab{\tau_{\rm R} q_2 \left(u\cdot p\right)}_0^+ + \frac{\beta\, m^2}{3} \ab{\frac{\tau_{\rm R} q_2}{\left(u\cdot p\right)}}_0^+ - \chi_a \ab{\tau_{\rm R} q_2}_0^- \right\} \left\{\ab{q_1 \left(u\cdot p\right)}_0^- - \mu \ab{q_1}_0^+ \right\}\right]
    \end{align}
    \begin{align}
        \mathcal{D}_\Pi^{\mu\nu} &= \frac{1}{\ab{q_4 \Big[\left(m^2/2\right) - \left(p\cdot \Delta \cdot p\right)/3 \Big]}_0^+} \bigg[ \ab{q_4 \Big[\left(u\cdot p\right)^2 - \left(p \cdot \Delta \cdot p\right)/3 \Big]}_0^+ u^{[\mu} \mathcal{C}_\Pi^{\nu]} + m^2 \ab{q_4\, \tau_{\rm R} s^{\mu\nu} s^{\alpha\beta}}_0^+ D_{\Pi,\, \alpha\beta} \nonumber\\
        &\qquad- m^2 \bigg\{\left(\chi_b - \frac{\beta}{3}\right) \ab{q_4\, \tau_{\rm R} \left(u\cdot p\right) s^{\mu\nu} s^{\alpha\beta}}_0^+ + \frac{\beta\, m^2}{3} \ab{\frac{q_4\, \tau_{\rm R}}{\left(u\cdot p\right)} s^{\mu\nu} s^{\alpha\beta}}_0^+ - \chi_a \ab{q_4\, \tau_{\rm R} s^{\mu\nu} s^{\alpha\beta}}_0^- \nonumber\\
        &\qquad+ \Big\{\ab{q_4\, \left(u\cdot p\right) s^{\mu\nu} s^{\alpha\beta}}_0^+ - \mu\ab{q_4\, s^{\mu\nu} s^{\alpha\beta}}_0^- \Big\} \frac{\mathcal{C}_2}{T^2} + \ab{q_4\, s^{\mu\nu} s^{\alpha\beta}}_0^- \frac{\mathcal{C}_3}{T} \bigg\} \omega_{\alpha\beta} \bigg],  \label{D_Pi-gen}\\
        \mathcal{D}_n^{\mu\nu\gamma} &= \frac{1}{\ab{q_4 \Big[\left(m^2/2\right) - \left(p\cdot \Delta \cdot p\right)/3 \Big]}_0^+} \bigg[ \ab{q_4 \Big[\left(u\cdot p\right)^2 - \left(p \cdot \Delta \cdot p\right)/3 \Big]}_0^+ u^{[\mu} \mathcal{C}_{n}^{\nu]\gamma} + m^2 D_{n,\,\alpha\beta}^{\qquad\gamma} \ab{q_4\, \tau_{\rm R} s^{\mu\nu} s^{\alpha\beta}}_0^+ \nonumber\\
        &\qquad- m^2 \bigg\{ \frac{1}{h_0} \!\ab{q_4\, \tau_{\rm R}\, p^{\ab{\gamma}} s^{\mu\nu} s^{\alpha\beta}}_0^+ \!\!- \ab{\frac{q_4\, \tau_{\rm R}}{\left(u\cdot p\right)} p^{\ab{\gamma}} s^{\mu\nu} s^{\alpha\beta}}_0^-\!\! - \beta^2\, \mathcal{C}_1 \ab{q_4\, p^{\ab{\gamma}} s^{\mu\nu} s^{\alpha\beta}}_0^+\! \bigg\} \omega_{\alpha\beta} \bigg], \label{D_n-gen}\\
        \mathcal{D}_\pi^{\mu\nu\alpha\beta} &= \frac{1}{\ab{q_4 \Big[\left(m^2/2\right) - \left(p\cdot \Delta \cdot p\right)/3 \Big]}_0^+} \bigg[ \ab{q_4 \Big[\left(u\cdot p\right)^2 - \left(p \cdot \Delta \cdot p\right)/3 \Big]}_0^+ u^{[\mu} \mathcal{C}_\pi^{\nu]\alpha\beta} + m^2 D_{\pi,\,\gamma\lambda}^{\quad~~\alpha\beta} \ab{q_4\, \tau_{\rm R} s^{\mu\nu} s^{\gamma\lambda}}_0^+ \nonumber\\
        &\qquad- m^2 \beta\, \omega_{\gamma\lambda} \ab{\frac{q_4\, \tau_{\rm R}}{\left(u\cdot p\right)} p^{\langle\alpha} p^{\beta\rangle} s^{\mu\nu} s^{\gamma\lambda}}_0^+ \bigg], \label{D_pi-gen}\\
        \mathcal{D}_\Sigma^{\mu\nu\gamma\alpha\beta} &= \frac{1}{\ab{q_4 \Big[\left(m^2/2\right) - \left(p\cdot \Delta \cdot p\right)/3 \Big]}_0^+} \bigg[ \ab{q_4 \Big[\left(u\cdot p\right)^2 - \left(p \cdot \Delta \cdot p\right)/3 \Big]}_0^+ u^{[\mu} \mathcal{C}_\Sigma^{\nu]\gamma\alpha\beta} + m^2 \ab{q_4\, \tau_{\rm R} s^{\mu\nu} s^{\phi\varphi}}_0^+ D_{\Sigma,\,\phi\varphi}^{\qquad\gamma\alpha\beta} \nonumber\\
        &\qquad+ m^2 \ab{\frac{q_4\, \tau_{\rm R}}{\left(u\cdot p\right)} p^{\ab{\gamma}} s^{\mu\nu} s^{\alpha\beta}}_0^+ \bigg] \label{D_Sig-gen}
    \end{align}
    where we have defined,
    \begin{align}
        \mathcal{C}_\Pi^{\nu} &= \frac{- m^2\, u_\mu}{\ab{\left(q_4/3\right) \left(p\cdot \Delta \cdot p\right)}_0^+} \bigg[\! \bigg\{\!\! \left(\chi_b \!-\! \frac{\beta}{3}\right)\! \ab{q_4\, \tau_{\rm R} \left(u\cdot p\right) s^{\mu\nu} s^{\alpha\beta}}_0^+ \!+ \frac{\beta\, m^2}{3} \!\ab{\!\frac{q_4\, \tau_{\rm R}}{\left(u\cdot p\right)} s^{\mu\nu} s^{\alpha\beta}\!}_0^+ \!\!\!-\! \chi_a \!\ab{q_4\, \tau_{\rm R} s^{\mu\nu} s^{\alpha\beta}}_0^- \nonumber\\
        &\qquad+ \Big\{\ab{q_4\, \left(u\cdot p\right) s^{\mu\nu} s^{\alpha\beta}}_0^+ - \mu\ab{q_4\, s^{\mu\nu} s^{\alpha\beta}}_0^- \Big\} \frac{\mathcal{C}_2}{T^2} + \ab{q_4\, s^{\mu\nu} s^{\alpha\beta}}_0^- \frac{\mathcal{C}_3}{T} \bigg\} \omega_{\alpha\beta} - \ab{q_4\, \tau_{\rm R} s^{\mu\nu} s^{\alpha\beta}}_0^+ D_{\Pi,\, \alpha\beta} \bigg] \label{C_Pi-gen}\\
        \mathcal{C}^{\nu\gamma}_n \!&=\! \frac{- m^2\, u_\mu}{\ab{\left(q_4/3\right) \left(p\cdot \Delta \cdot p\right)}_0^+} \bigg[\! \bigg\{\! \frac{1}{h_0} \!\ab{\!q_4\, \tau_{\rm R}\, p^{\ab{\gamma}} s^{\mu\nu} s^{\alpha\beta}\!}_0^+ \!\!-\! \ab{\!\frac{q_4\, \tau_{\rm R}}{\left(u\cdot p\right)} p^{\ab{\gamma}} s^{\mu\nu} s^{\alpha\beta}\!}_0^- \!\!\!-\! \beta^2\, \mathcal{C}_1 \!\ab{\!q_4\, p^{\ab{\gamma}} s^{\mu\nu} s^{\alpha\beta}\!}_0^+\!\!\bigg\} \omega_{\alpha\beta} \nonumber\\
        &\qquad-\! D_{n,\, \alpha\beta}^{\qquad\!\gamma} \ab{q_4\, \tau_{\rm R}\, s^{\mu\nu} s^{\alpha\beta}}_0^+ \!\bigg] \label{C_n-gen}\\
        \mathcal{C}^{\nu\gamma\lambda}_\pi &= \frac{- m^2\, u_\mu}{\ab{\left(q_4/3\right) \left(p\cdot \Delta \cdot p\right)}_0^+} \bigg[ \beta\, \omega_{\alpha\beta} \ab{\frac{q_4\, \tau_{\rm R}}{\left(u\cdot p\right)} p^{\langle\gamma} p^{\lambda\rangle} s^{\mu\nu} s^{\alpha\beta}}_0^+ - D_{\pi,\,\alpha\beta}^{\qquad\gamma\lambda} \ab{q_4\, \tau_{\rm R} s^{\mu\nu} s^{\alpha\beta}}_0^+ \bigg] \label{C_pi-gen}\\
        \mathcal{C}^{\nu\gamma\phi\varphi}_\Sigma &= \frac{- m^2\, u_\mu}{\ab{\left(q_4/3\right) \left(p\cdot \Delta \cdot p\right)}_0^+} \bigg[ \ab{\frac{q_4\, \tau_{\rm R}}{\left(u\cdot p\right)} p^{\ab{\gamma}} s^{\mu\nu} s^{\phi\varphi}}_0^+ + \ab{q_4\, \tau_{\rm R} s^{\mu\nu} s^{\alpha\beta}}_0^+ D_{\Sigma,\,\alpha\beta}^{\qquad\gamma\phi\varphi} \bigg]. \label{C_Sig-gen}
    \end{align}

The coefficients defined in Eqs.~\eqref{D-b,T,m} are given as,
\begin{align}
    \chi_a = \beta \left(\frac{I_{30}^+ I_{21}^- - I_{31}^+ I_{20}^-}{I_{30}^+ I_{10}^+ - I_{20}^- I_{20}^-}\right);
    \qquad\qquad
    \chi_b = \beta \left(\frac{I_{20}^- I_{21}^- - I_{31}^+ I_{10}^+}{I_{30}^+ I_{10}^+ - I_{20}^- I_{20}^-}\right)
\end{align}

The coefficients defined in Eqs.~\eqref{del_n,e,P-NS}-\eqref{del_S^lmn-NS} are given as,
\begin{align}
    \nu &= 2 \bigg[ \!\left(\chi_b - \frac{\beta}{3}\right) \!\ab{\tau_{\rm R} \left(u\cdot p\right)^2}_0^- \!+ \frac{\beta\, m^2}{3} \ab{\tau_{\rm R}}_0^- - \chi_a \ab{\tau_{\rm R} \left(u\cdot p\right)}_0^+ \nonumber\\
    &\qquad+ \frac{\mathcal{C}_2}{T^2} \left(\ab{\left(u\cdot p\right)^2}_0^- - \mu \ab{\left(u\cdot p\right)}_0^+\right) + \frac{\mathcal{C}_3}{T} \ab{\left(u\cdot p\right)}_0^- \!\bigg], \\
    e &= 2 \bigg[ \!\left(\chi_b - \frac{\beta}{3}\right) \!\ab{\tau_{\rm R} \left(u\cdot p\right)^3}_0^+ \!+ \frac{\beta\, m^2}{3} \ab{\tau_{\rm R} \left(u\cdot p\right)}_0^+ - \chi_a \ab{\tau_{\rm R} \left(u\cdot p\right)^2}_0^- \nonumber\\
    &\qquad+ \frac{\mathcal{C}_2}{T^2} \ab{\left(u\cdot p\right)^3}_0^+ + \frac{\mathcal{C}_3}{T} \ab{\left(u\cdot p\right)^2}_0^+ - \frac{\mu\, \mathcal{C}_2}{T^2} \ab{\left(u\cdot p\right)^2}_0^- \!\bigg], \\
    \rho &= - \frac{2}{3} \bigg[ \!\left(\chi_b - \frac{\beta}{3}\right) \!\ab{\tau_{\rm R} \left(u\cdot p\right) \left(p\cdot\Delta\cdot p\right)}_0^+ \!+ \frac{\beta\, m^2}{3} \ab{\frac{\tau_{\rm R} \left(p\cdot\Delta\cdot p\right)}{\left(u\cdot p\right)}}_0^+ - \chi_a \ab{\tau_{\rm R} \left(p\cdot\Delta\cdot p\right)}_0^- \nonumber\\
    &\qquad+ \frac{\mathcal{C}_2}{T^2} \ab{\left(u\cdot p\right) \left(p\cdot\Delta\cdot p\right)}_0^+ + \frac{\mathcal{C}_3}{T} \ab{\left(p\cdot\Delta\cdot p\right)}_0^+ - \frac{\mu\, \mathcal{C}_2}{T^2} \ab{\left(p\cdot\Delta\cdot p\right)}_0^- \!\bigg], \\
    \kappa_n^{\mu\alpha} &= 2 \left[\frac{1}{h_0} \ab{\tau_{\rm R} p^{\ab{\mu}} p^{\ab{\alpha}}}_0^- - \frac{\mathcal{C}_1}{T^2} \ab{p^{\ab{\mu}} p^{\ab{\alpha}}}_0^- - \ab{\frac{\tau_{\rm R} p^{\ab{\mu}} p^{\ab{\alpha}}}{\left(u\cdot p\right)}}_0^+ \right], \\
    \kappa_h^{\mu\alpha} &= 2 \left[\frac{1}{h_0} \ab{\tau_{\rm R} \left(u\cdot p\right) p^{\ab{\mu}} p^{\ab{\alpha}}}_0^+ - \frac{\mathcal{C}_1}{T^2} \ab{\left(u\cdot p\right) p^{\ab{\mu}} p^{\ab{\alpha}}}_0^+ - \ab{\tau_{\rm R} p^{\ab{\mu}} p^{\ab{\alpha}}}_0^- \right], \\
    \eta^{\mu\nu\alpha\beta} &= \beta \ab{\frac{\tau_{\rm R} p^{\langle\mu} p^{\nu\rangle} p^{\langle\alpha} p^{\beta\rangle}}{\left(u\cdot p\right)}}_0^+,
\end{align}
\begin{align}
    B_\Pi^{\lambda\mu\nu} &= \frac{1}{2} \bigg[ \bigg\{\left(\chi_b - \frac{\beta}{3}\right) \ab{\tau_{\rm R} \left(u\cdot p\right) p^\lambda s^{\mu\nu} s^{\alpha\beta}}_0^+ + \frac{\beta\, m^2}{3} \ab{\frac{\tau_{\rm R} p^\lambda s^{\mu\nu} s^{\alpha\beta}}{\left(u\cdot p\right)}}_0^+ - \chi_a \ab{\tau_{\rm R} p^\lambda s^{\mu\nu} s^{\alpha\beta}}_0^- + \frac{\mathcal{C}_2}{T^2} \ab{\left(u\cdot p\right) p^\lambda s^{\mu\nu} s^{\alpha\beta}}_0^+ \nonumber\\
    &\qquad+ \frac{\mathcal{C}_3}{T} \ab{p^\lambda s^{\mu\nu} s^{\alpha\beta}}_0^+ - \frac{\mu\, \mathcal{C}_2}{T^2} \ab{p^\lambda s^{\mu\nu} s^{\alpha\beta}}_0^-\bigg\} \omega_{\alpha\beta} - \ab{\tau_{\rm R} p^\lambda s^{\mu\nu} s^{\alpha\beta}}_0^+ D_{\Pi,\alpha\beta} + \ab{p^\lambda s^{\mu\nu} s^{\alpha\beta}}_0^+ \mathcal{D}_{\Pi,\alpha\beta} \bigg]\\
    B_n^{\lambda\mu\nu\alpha} &= \frac{1}{2} \!\bigg[\!\left\{\! \frac{1}{h_0} \!\ab{\!\tau_{\rm R}\, p^{\ab{\alpha}} p^\lambda s^{\mu\nu} s^{\beta\gamma}}_0^+ \!\!\!- \frac{\mathcal{C}_1}{T^2} \ab{\!p^{\ab{\alpha}} p^\lambda s^{\mu\nu} s^{\beta\gamma}}_0^+ \!\!\!- \ab{\!\frac{\tau_{\rm R} p^{\ab{\alpha}} p^\lambda s^{\mu\nu} s^{\beta\gamma}}{\left(u\cdot p\right)}}_0^- \!\right\} \omega_{\beta\gamma} \nonumber\\
    &\qquad- \ab{\tau_{\rm R}\, p^\lambda s^{\mu\nu} s_{\beta\gamma}}_0^+ D_n^{\beta\gamma\alpha} + \ab{p^\lambda s^{\mu\nu} s_{\beta\gamma}}_0^+ \mathcal{D}_n^{\beta\gamma\alpha} \!\bigg]\\
    B_\pi^{\lambda\mu\nu\alpha\beta} &= \frac{\beta}{2} \left[\ab{\frac{\tau_{\rm R}\, p^{\langle\alpha} p^{\beta\rangle} p^\lambda s^{\mu\nu} s^{\gamma\varphi}}{\left(u\cdot p\right)}}_0^+ \omega_{\gamma\varphi} - \ab{\tau_{\rm R}\, p^\lambda s^{\mu\nu} s_{\gamma\varphi}}_0^+ D_\pi^{\gamma\varphi\alpha\beta} + \ab{p^\lambda s^{\mu\nu} s_{\gamma\varphi}}_0^+ \mathcal{D}_\pi^{\gamma\varphi\alpha\beta} \right]\\
    B_\Sigma^{\lambda\mu\nu\gamma\alpha\beta} &= \frac{1}{2} \left[- \ab{\frac{\tau_{\rm R}}{\left(u\cdot p\right)} p^{\ab{\gamma}} p^\lambda s^{\mu\nu} s^{\alpha\beta}}_0^+ - \ab{\tau_{\rm R} p^\lambda s^{\mu\nu} s_{\varphi\rho}}_0^+ D_{\Sigma}^{\varphi\rho\gamma\alpha\beta} + \ab{p^\lambda s^{\mu\nu} s_{\varphi\rho}}_0^+ \mathcal{D}_{\Sigma}^{\varphi\rho\gamma\alpha\beta}\right]
\end{align}
\end{widetext}

\bibliography{ref}{}

\end{document}